\documentclass[a4paper,12pt]{article}
\usepackage{amsmath,amssymb,algorithmic}
\usepackage{latexsym}
\usepackage[T1]{fontenc}
\usepackage[utf8]{inputenc}
\usepackage{textcomp}
\usepackage{times}
\usepackage[pdftex]{graphicx}
\usepackage[authoryear,square]{natbib}
\usepackage[frenchb]{babel}
\usepackage{babelbib}
\usepackage{hyperref}

\hypersetup{
colorlinks=true, 
citecolor=blue, %
pdftoolbar=false,
bookmarksopen=true,
breaklinks=true, 
urlcolor=blue, 
linkcolor=blue, 
}

\def\gl{\guillemotleft }
\def\gr{\guillemotright }

\newcommand{\bm}[1]{\mathbf{ #1}}

\author{Philippe Besse\thanks{Universit\'e de Toulouse -- INSA, Institut de Math\'ematiques, UMR CNRS 5219} \and Brendan Guillouet\thanks{Institut de Math\'ematiques, UMR CNRS 5219} \and Jean-Michel Loubes\thanks{Universit\'e de Toulouse -- UT3, Institut de Math\'ematiques, UMR CNRS 5219}}
\title{\sc Apprentissage sur Donn\'ees Massives \\ trois cas d'usage avec \\ R, Python et Spark} 

\DeclareGraphicsExtensions{.pdf,.jpg,.png}

\begin{document}
\sloppy
\maketitle

\begin{quote}
{\bf R\'esum\'e}: 
La gestion et l'analyse de donn\'ees massives sont syst\'ematiquement associ\'ees \`a une architecture de donn\'ees distribu\'ees dans des environnements {\it Hadoop} et maintenant {\it Spark}. Cet article propose aux statisticiens une introduction \`a ces technologies en comparant les performances obtenues par l'utilisation \'el\'ementaire de trois environnements de r\'ef\'erence: R, Python {\it Scikit-learn}, Spark {\it MLlib} sur trois cas d'usage publics: reconnaissance de caract\`eres, recommandation de films, cat\'egorisation de produits. Comme principal r\'esultat, il en ressort que si Spark est tr\`es performant pour la pr\'eparation des donn\'ees et la recommandation par filtrage collaboratif  (factorisation non n\'egative), les impl\'ementations actuelles des m\'ethodes classiques d'apprentissage (r\'egression logistique, for\^ets al\'eatoires) dans {\it MLlib} ou {\it SparkML} ne concurrencent pas ou mal une utilisation habituelle de ces m\'ethodes (R, Python Scikit-learn) dans une architecture int\'egr\'ee au sens de non distribu\'ee.

{\bf Mots-clefs}: Science des Donn\'ees; Apprentissage Machine; Statistique; Analyse de Donn\'ees Massives; Cas d'Usage.

{\bf Abstract}: 
Management and analysis of big data are systematically associated with a data distributed architecture in the {\it Hadoop}  and now {\it Spark} frameworks. This article offers an introduction for statisticians to these technologies by comparing the performance obtained by the direct use of three reference environments: R, Python {\it Scikit-learn}, Spark {\it MLlib} on three public use cases: character recognition, recommending films, categorizing products. As main result, it appears that, if Spark is very efficient for data munging and recommendation by collaborative filtering (non-negative factorization), current implementations of conventional learning methods (logistic regression, random forests) in {\it MLlib} or {\it SparkML} do not ou poorly compete habitual use of these methods (R, Python Scikit-learn) in an integrated or undistributed architecture.

{\bf Keywords}: Data Science; Machine Learning; Statistics; Big Data Analytics; Use Cases.
\end{quote}




\section{Introduction}
\subsection{Objectif}
L'objectif de cet article est d'ouvrir une réflexion sur le choix des meilleures mises en \oe uvre de techniques d'apprentissage face à des données massives. Cet objectif ambitieux se heurte à de nombreuses difficultés : il présuppose une définition de ce que sont des \emph{données massives}, donc les spécificités de l'environnement de travail et le choix de critères de comparaison.

De façon triviale, des données deviennent massives lorsqu'elles excèdent la capacité de la mémoire vive (RAM) de l'ordinateur (quelques giga-octets), puis réellement massives lorsqu'elles excèdent celle du disque dur (quelques tera-octets) et doivent être distribuées sur plusieurs disques voire machines. Cette définition dépend évidemment de l'environnement matériel mais les principaux problèmes méthodologiques et algorithmiques sont soulevés lorsque les données sont effectivement distribuées sur plusieurs machines.  Le principal objectif est alors d'éviter des transferts coûteux de données entre ordinateurs ou même entre disques et unités centrales ; analyse et calculs sont déportés sur le lieu de stockage et doivent se limiter à une seule lecture des données.

L'environnement de travail peut-être un poste de travail personnel avec plus ou moins de mémoire, de processeurs, de cartes graphiques (GPU), l'accès à un puissant serveur de calcul intensif ou encore, c'est la caractéristique originale de la prise en compte de données massives, à un \emph{cluster} de calcul réel ou virtuel, privé ou loué. Pouvoir transférer les calculs d'un poste personnel utilisé pour la réalisation d'un prototype, au déploiement, passage à l'échelle, d'un grand {\it cluster} à l'aide du même code est un élément essentiel d'efficacité car le coût de développement humain l'emporte largement sur celui de location d'un espace de calcul.

Les évaluations et comparaisons classiques de performances en apprentissage sur des données de faible taille se compliquent encore avec l'explosion combinatoire du nombre d'implémentations des algorithmiques ainsi qu'avec la diversité des technologies de parallélisation et celle des architectures matérielles. Le nombre de paramètres à considérer rend ces comparaisons beaucoup trop complexes ou soumises à des choix bien arbitraires. Face à cette complexité, l'objectif est nécessairement réduit. Loin de pouvoir discuter la réelle optimalité des technologies d'analyse de données massives, on se propose d'aborder les intérêts respectifs de quelques stratégies autour de trois cas d'usage  devenus classiques afin d'illustrer, ne serait-ce que la complexité des problèmes et la pertinence de certains choix.

L'objectif est donc réduit à la comparaison des implémentations, dans trois environnements très répandus ou en pleine expansion : R, Python ({\it Scikit-learn}) et {\it Spark-MLlib}, de trois méthodes parmi les plus utilisées : les forêts aléatoires (\citet{brei-2001}) en classification supervisée, la factorisation non négative de matrices (NMF) pour apprendre des recommandations et enfin la régression logistique faisant suite au pré-traitement d'un corpus volumineux de textes. 

D'autres approches, d'autres technologies, d'autres méthodes, certes importantes et faisant par exemple appel à l'apprentissage profond (\emph{deep learning}) pour l'analyse d'images, à des méthodes d'optimisation stochastique ou de décision séquentielle... nécessiteraient des développements spécifiques conséquents ; elles sont volontairement laissées de côté. 

Il s'agit simplement d'illustrer, sur des cas d'usage, le choix qui se pose au statisticien de passer, ou non, à un environnement technologique  plus complexe : Python {\it Scikit-Learn} (\citet{pedr-2011}), {\it Spark} (\citet{zaha-2012}) que celui : R (\citet{team-2016}) qu'il maîtrise depuis de longues années.

Tous les scripts des programmes utilisés en R, Python et Pyspark sont disponibles sur le site {\tt github/wikistat.fr} sous la forme de {\it notebooks} ou calepins {\it Jupyter}.

Les auteurs tiennent à remercier l'équipe de la société {\it Hupi} de Bidart et son directeur M. Moréno qui ont mis à notre disposition la plateforme matérielle ({\it cluster}) utilisée pour aborder l'environnement {\it Spark Hadoop} et qui ont également assuré un support efficace pour exploiter au mieux les technologies pour données distribuées.

\subsection{Nouvelle Science des Données}
La fouille de données (\emph{data mining}) des années 90 a largement promu de nouveaux métiers et surtout de nouvelles suites logicielles commerciales intégrant gestion, transformations et analyse statistique des données pour s'adresser à de plus vastes marchés, principalement le marketing. L'objectif était déjà la valorisation (gestion de la relation client) de données acquises par ailleurs et sans planification expérimentale spécifique à une démarche statistique traditionnelle. Une deuxième étape, en lien avec l'accroissement du volume et la complexité des données,  fut l'explosion de leur très grande dimension avec la multitude des omiques (génome, transcriptome, protéome...) produite par les technologies de séquençage. Ce fut l'expansion de la Bioinformatique et, pour le statisticien confronté au fléau de la dimension, la nécessaire recherche de modèles parcimonieux. Une troisième étape est liée au développement d'internet, du commerce en ligne et des réseaux sociaux. Ce sont les principales sources de production et d'analyse de données massives (\emph{big data analytics}).  Comme pour la fouille des données des années 90, mais avec une ampleur sans commune mesure, l'histoire se répète avec l'explosion du marché des espaces publicitaires ({\it advertasing}) en ligne ({\it Google, Facebook...}) et celles des services ({\it Amazon Web Service...}) associés.

\citet{frie-1997} soulevait déjà la question : {\it Is data mining an intellectual discipline?} Les enjeux actuels font \emph{changer d'échelle} pour s'interroger sur la naissance plus prestigieuse d'une \gl~nouvelle\gr~ \emph{Science, des Données}, accompagnée d'un considérable battage médiatique. Celui-ci glorifie les exploits de la ruée vers le nouvel eldorado des investissements et emplois mais stigmatise aussi les risques éthiques, juridiques ou les fiascos annoncés. Cette évolution, tant historique que méthodologique, et son influence sur les programmes académiques est décrite par \citet{bess-2016}. Elle n'est pas reprise ici mais nous en retenons le principal élément. Celui-ci s'apparente à l'émergence d'un nouveau paradigme provoquée par les changements d'échelles de volume, variété, vélocité des données qui  bousculent  les pratiques, tant en Statistique qu'en Apprentissage Machine. 

L'estimation d'un modèle prédictif, qu'il soit statistique ou d'apprentissage supervisé, est la recherche d'un compromis optimal entre biais ({\emph{erreur d'approximation}) et variance ({\emph{erreur d'estimation}) donc  d'un modèle parcimonieux évitant le sur-ajustement. \`A cette question centrale en Apprentissage, vient s'ajouter un nouveau problème  d'{\emph{Optimisation} sous la forme d'un troisième terme d'erreur. Les ressources (mémoire, temps) sont contraintes; comment minimiser le terme d'erreur d'optimisation dû, soit à un sous-échantillonnage pour satisfaire aux contraintes de mémoire, soit à la limitation des calculs ou méthodes utilisables sur une base d'apprentissage très volumineuse et distribuée ? Une fonction objectif globale à minimiser pourrait être définie à la condition d'évaluer le coût induit par des erreurs de prévision dues à l'échantillonnage, à mettre en balance avec le coût de l'estimation avec données massives sur des serveurs les supportant. Le problème général étant bien trop complexe, nous nous restreignons à des comparaisons plus qualitatives.

\subsection{Contenu}
La section 2 introduit l'environnement technologique de référence \emph{Spark} (\citet{zaha-2012}) associé à la gestion de données massive {\it Hadoop distributed file system}. L'importance de la notion de \emph{résilience} des bases de données est illustrée par l'exemple de l'algorithme élémentaire de \citet{forg-1965} modifié pour satisfaire aux contraintes des fonctionnalités {\it MapReduce} d'{\it Hadoop}. Sont introduites également les librairies {\it MLlib, SparkML} développées pour exécuter les méthodes de modélisation et apprentissage dans ce contexte. 

Les sections suivantes abordent trois cas d'usage sur des données publiques ou rendues publiques : reconnaissance de caractères (MNIST), recommandation de films (MovieLens), catégorisation de produits (Cdiscount) avec l'objectif de comparer les performances (temps, précision) des environnements R, Python et Spark sur ces exemples de taille raisonnable.

Les comparaisons ont été opérées sur des machines de faible taille / coût : Lenovo X240 (Windows 7) avec processeur 4 c\oe urs cadencé à 1,7 GHz et 8Go de RAM, MacBook Pro (OS X Yosemite) avec processeur 4 c\oe urs cadencé à 2,2 GHz et 16Go de RAM, cluster (Linux, Spark) avec au plus 1 n\oe ud maître et 8 n\oe uds excécuteurs de 7Go de RAM.  Ces configurations ne sont pas très réalistes face à des données massives. Néanmoins les résultats obtenus, sont suffisamment explicites pour illustrer les contraintes du compromis taille des données {\it vs.} précision pour une taille de mémoire fixée.

La dernière section conclut sur l'opportunité des choix en présence.

\section{Principaux environnements technologiques}
Les choix de méthodes offertes par les librairies et les performances d'une analyse dépendent directement de l'environnement matériel et logiciel utilisé. La profusion des possibilités rendent ces choix difficiles. Il serait vain de chercher à en faire une synthèse, mais voici quelques éléments de comparaison parmi les environnements les plus pratiqués ou au mois les plus médiatisés et accessibles car sous la licence de l'{\it Apache Software Fondation}.

\subsection{Nouveau modèle économique}
Rappelons tout d'abord que le déluge des données, conséquence de la \emph{datafication} de notre quotidien, entraîne un profond changement des modèles économiques en lien avec l'analyse de ces données. Changement qui impacte directement les développements et donc disponibilités des environnements accessibles aux entreprises et aux établissements académiques. Les équipements trop chers sont loués, les langages et logiciels sont libres mais les concepts et méthodes complexes à assimiler et mettre en \oe uvre sont des services (formations, plateformes) monnayables. Plus généralement, tout un ensemble de ces services et une nomenclature associée se développent avec l'industrialisation, la commercialisation du \emph{cloud computing} : \emph{software as a service} (SaaS), \emph{infrastructure as a service} (IaaS), \emph{platform as a service} (PaaS), \emph{desktop as a service} (DaaS)... 

\`A titre d'illustration, citons seulement quelques entreprises surfant sur la vague des nouvelles  technologies : \emph{Enthought} (Canopy) et \emph{Continuum analytics} (Anaconda) proposent des distributions libres de Python et, c'est important, faciles à installer ainsi que des environnements plus élaborés payants et de la formation associée. \emph{Horthonworks} et \emph{Cloudera} diffusent des environnements de {\it Hadoop} incluant {\it Spark}. Les créateurs (\citet{zaha-2012}) de \emph{Spark} ont fondé \emph{databricks} : \emph{Data science made easy, from ingest to production}, pour principalement vendre de la formation et une certification. Trevor Hastie et Ron Tibshirani conseillent \emph{Oxdata} qui développe (\emph{H20}) avec notamment une forme d'interface entre R et le système \emph{Hadoop}.

Dans le même mouvement ou cherchant à ne pas perdre trop de terrain par rapport à {\it Amazon Web Services}, les grands éditeurs ou constructeurs intègrent, avec plus ou moins de succès, une offre de service à leur modèle économique d'origine : {\it Google Cloud platform, IBM Analytics, Microsoft Azure, SAS Advanced Analytics...}

\subsection{\it Hadoop, MapReduce, Spark}
\emph{Hadoop} ({\it distributed file system}) est devenu la technologie systématiquement associée à la notion de données considérées comme massives car distribuées. Largement développé par \emph{Google} avant de devenir un projet de la fondation {\it Apache}, cette technologie répond à des besoins spécifiques de centres de données : stocker des volumétries considérables en empilant des milliers de cartes d'ordinateurs et disques de faible coût, plutôt que de mettre en \oe uvre un supercalculateur, tout en préservant la fiabilité par une forte tolérance aux pannes. Les données sont dupliquées et, en cas de défaillance, le traitement est poursuivi, une carte remplacée, les données automatiquement reconstruites, sans avoir à arrêter le système. 

Ce type d'architecture génère en revanche un coût algorithmique. Les n\oe uds de ces serveurs ne peuvent communiquer que par couples (clef, valeur) et les différentes étapes d'un traitement doivent pouvoir être décomposées en étapes fonctionnelles élémentaires comme celles dites de \emph{MapReduce}. Pour des opérations simples, par exemple de dénombrement de mots, d'adresses URL, des statistiques élémentaires, cette architecture s'avère efficace. Une étape {\it Map} réalise, en parallèle, les dénombrements à chaque n\oe ud ou exécuteur (\emph{workers}) d'un \emph{cluster} d'ordinateurs, le résultat est un ensemble de couples : une clef (le mot, l'URL à dénombrer) associée à un résultat partiel. Les clefs identiques sont regroupées dans une étape intermédiaire de tri (\emph{shuffle}) au sein d'une même étape \emph{Reduce} qui fournit pour chaque clef le résultat final.

Dans cette architecture, les algorithmes sont dits \emph{échelonnables} de l'anglais \emph{scalable} si le temps d'exécution décroît linéairement avec le nombre d'exécuteurs dédiés au calcul. C'est immédiat pour des dénombrements, des calculs de moyennes, ce n'est pas nécessairement le cas pour des algorithmes itératifs complexes.  La méthode des $k$ plus proches voisins n'est pas échelonnable au contraire des algorithmes de classification non-supervisée par réallocation dynamique (\emph{e.g.} Forgy, $k$-means) qui opèrent par itérations d'étapes \emph{MapReduce}.

L'exemple de l'algorithme de \citet{forg-1965} est très révélateur.
\begin{quote}
\begin{itemize}
	\item {\bf Initialisation} de l'algorithme par définition d'une fonction de distance et désignation aléatoire de $k$ centres.
	\item {\bf Jusqu'à} convergence :
	\begin{itemize}
	\item L'étape {\bf Map} calcule, en parallèle, les distances de chaque observation aux $k$ centres courants. Chaque observation (vecteur de valeurs) est affectée au centre (clef) le plus proche. Les couples : (clef ou numéro de centre, vecteur des valeurs) sont communiqués à l'étape \emph{Reduce}.
	\item Une étape intermédiaire implicite {\bf Shuffle} adresse les couples de même clef à la même étape suivante.
	\item {\bf Pour} chaque clef désignant un groupe, l'étape {\bf Reduce} calcule les nouveaux barycentres, moyennes des valeurs des variables des individus partageant la même classe c'est-à-dire la même valeur de clef.
	\end{itemize}
\end{itemize}
\end{quote}

Principal problème de cette implémentation, le temps d'exécution économisé par la parallélisation des calculs est fortement pénalisé  par la nécessité d'écrire et relire toutes les données entre deux itérations. 
\subsection{\it Spark}
C'est la principale  motivation du développement de la technologie \emph{Spark} (\citet{zaha-2012}) à l'université de Berkeley. Cette couche logicielle au-dessus de systèmes de gestion de fichiers comme \emph{Hadoop} introduit la notion de base de données \emph{résiliente} (\emph{resilient distributed dataset} ou RDD) dont chaque partition reste, si nécessaire, présente en mémoire entre deux itérations pour éviter réécriture et relecture. Cela répond bien aux principales contraintes : des données massives ne doivent pas être déplacées et un résultat doit être obtenu par une seule opération de lecture. 

Techniquement, ces RDDs sont manipulées par des commandes en langage \emph{Java} ou \emph{Scala} mais il existe des API (\emph{application programming interface}) acceptant des commandes en Python (\emph{PySpark}) et en R. {\it Spark} intègre beaucoup de fonctionnalités réparties en quatre modules : \emph{GRAPHX} pour l'analyse de graphes ou réseaux, \emph{streaming} pour le traitement et l'analyse des flux, {\it SparkSQL} pour l'interrogation et la gestion de bases de tous types  et la librairie {\it MLlib} pour les principaux algorithmes d'apprentissage. En plein développement, cet environnement comporte (version 1.6) des incohérences. {\it SparkSQL} génère et gère une nouvelle classe de données {\it DataFrame} (similaire à R) mais qui n'est pas connue de {\it MLlib} qui va progressivement être remplacée par {\it SparkML} dans les versions à venir...

Dans la présentation qui en est faite, {\it Spark} apparaît donc comme un cadre général ({\it framework})  permettant de connecter la plupart des technologies développées sous forme de projet \emph{Apache}. Tous types de fichier (JSON, csv, RDDs...) et types de données structurées ou non, tous types de flux de données (objets connectés, tweets, courriels...) peuvent ainsi être gérés, agrégés par des opérations classiques de sélection, fusion à l'aide de commandes utilisant une syntaxe dérivée de SQL ({\it structured querry language}) avant d'être utilisés pour des modélisations.

\subsection{Librairie {\it MLlib}}
Cet article est focalisé sur l'utilisation de quelques méthodes de transformation et modélisation sous  {\it Spark} dont celles de {\it MLlib} décrites par \citet{pent-2015}. Cette librairie regroupe les algorithmes d'apprentissage adaptés à des bases de données résilientes et qui supportent donc le \emph{passage à l'échelle} du volume des données. Un programme mis au point sur de petits échantillons et un poste de travail personnel peut en principe s'exécuter en l'état sur un \emph{cluster} de calcul (\emph{e.g. Amazon Web Service}) pour analyser des données massives même si la \emph{scalabilité} n'est pas strictement atteinte.

Pour cette librairie en plein développement et aussi en migration vers {\it SparkML}, seule la documentation en ligne, malheureusement fort succincte, est à jour concernant la liste des méthodes disponibles et leurs options d'utilisation. En voici un rapide aperçu (version 1.6) :
\begin{description}
\item[\it Statistique de base] : Univariée, corrélation, échantillonnage stratifié, tests d'hypothèse, générateurs aléatoires, transformations (standardisation, quantification de textes avec hashage et TF-IDF pour  {\it vectorisation}), sélection ($\chi$2) de variables (\emph{features}).
\item[\it Exploration multidimensionnelle] Classification non-supervisée ($k$-means avec version en ligne, modèles de mélanges gaussiens, LDA (\emph{Latent Dirichlet Allocation}), réduction de dimension (SVD et ACP mais en Java ou Scala pas en Python), factorisation non négative de matrice (NMF) par moindres carrés alternés (ALS).
\item[\it Apprentissage] Méthodes linéaires : SVM, régression gaussienne et binomiale ou logistique avec pénalisation L1 ou L2 ; estimation par gradient stochastique, ou L-BFGS ; classifieur bayésien naïf, arbre de décision, forêts aléatoires, boosting (\emph{gradient boosting machine} en Scala).
\end{description}

Les sections qui suivent proposent une étude comparative des environnements R, Python ({\it Scikit-Learn}) et {\it Spark  MLlib} sur trois cas d'usage de données publiques. Les données ne sont pas excessivement volumineuses mais sont considérées comme telles en mettant en \oe uvre des technologies d'algorithmes distribués capables en principe de passer sans modification à des échelles plus importantes. Elles le sont pas ailleurs suffisamment pour mettre en défaut des implémentations pas ou mal optimisées comme la librairie {\it randomForest} de R.

\section{Reconnaissance de caractères (MNIST)}

La reconnaissance de caractères manuscrits, notamment les chiffres de codes postaux, est un vieux problème de classification supervisée qui sert depuis de nombreuses années de base de comparaison entre les méthodes (cf. figure~\ref{mnist}). Le site  \emph{MNIST DataBase} fournit des données (\citet{lecu-1998}) et une liste de références, de 1998 à 2012 proposant des stratégies d'analyse avec des taux d'erreur de 12\% à 0,23\%. Les méthodes les plus récentes s'affrontent toujours sur ce type de données : \citet{brun-2013} ... \citet{lee-2016} ainsi que dans un concours \emph{Kaggle} se terminant en décembre 2016. 

\begin{figure}
\centerline{\includegraphics[width=7cm]{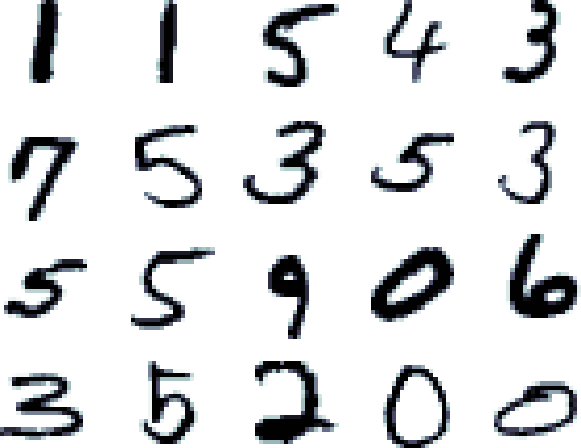}}
\caption{\it Quelques exemples d'images de caractères de la base MNIST. }\label{mnist}
\end{figure}

De façon très schématique, plusieurs stratégies sont développées dans une vaste littérature sur ces données.  
\begin{itemize}
\item Utiliser une méthode classique : $k$ plus proches voisins, forêt aléatoire... sans trop raffiner mais avec des temps d'apprentissage rapides conduit à un taux d'erreur autour de 3\%. 
\item Ajouter  ou intégrer un pré-traitement des données permettant de recaler les images par des distorsions plus ou moins complexes.
\item Construire une mesure de distance adaptée au problème car invariante par rotation, translation, homothétie,  puis l'intégrer dans une technique d'apprentissage adaptée comme les $k$ plus proches voisins. \citet{sima-1998} définissent une distance dite tangentielle entre les images de caractères possédant ces propriétés d'invariance.
\item D'autres pistes peuvent être explorées notamment celles des machines à noyaux par \citet{mull-2001}  qui illustrent ce type d'apprentissage par les données de MNIST sans semble-t-il faire mieux que la distance tangentielle. 
\item Les réseaux de neurones, renommés apprentissage profond (\emph{deep learning}), et implémentant une architecture modélisant une convolution (\emph{convolutional neural network}) impliquant les propriétés recherchées d'invariance, sont les algorithmes les plus compétitifs et les plus comparés (\citet{wan-2013}). Noter le coût de calcul assez prohibitif pour l'estimation et l'optimisation de ces architectures complexes qui nécessitent des moyens matériels et des librairies sophistiquées reprenant toutes ce même jeu de données dans leur tutoriel. C'est le cas de {\it torch} en langage {\it Lua} ou de \emph{Lasagne} (Theano) en Python. Même chose pour {\it H2O}, dont le tutoriel, très commercial, optimise les paramètres à partir de l'échantillon test malgré un risque évident de sur-apprentissage, ou encore pour \emph{Tensor Flow} rendu récemment (12-2015)  accessible par {\it Google}. 
\item Toujours en connexion avec le \emph{deep learning} et illustrée par les mêmes données, \citet{brun-2013} utilisent une décomposition, invariante par transformations, des images sur des bases d'ondelettes par \emph{scattering}. 
\item ...
\end{itemize}
Attention, l'objectif de cette section n'est pas de concourir avec les meilleures solutions publiées. Les données sont simplement utilisées pour comparer les performances des implémentations de solutions élémentaires dans l'objectif d'un passage à l'échelle volume.
\subsection{Les données}
Les données représentent 60 000 images en niveaux de gris de chiffres manuscrits écrits  par plusieurs centaines de personnes sur une tablette. Un ensemble de pré-traitements restreint les dimensions des images à $28 \times 28 = 784$ pixels. Celles-ci ont ensuite été normalisées par des transformations élémentaires. L' échantillon test indépendant est constitué de la même manière de 10 000 images. 

Ces données ne sont pas réellement massives, elles tiennent en mémoire, mais leur volume suffit à montrer les limites, notamment en temps de calcul ou occupation mémoire, de certains algorithmes.

\subsection{Solutions}
Trois environnements sont comparés : R, Python ({\it Scikit-learn}) et {\it Spark} ({\it MLlib}). Une simple  classification non supervisée ($k$-means) a également été testée sur les données. Les résultats sont équivalents entre les environnements à quelques remarques près. L'implémentation de R signale une difficulté à converger mais fournit avec le même temps d'exécution des résultats similaires à ceux de {\it Scikit-learn}. En revanche, l'algorithme des $k$ plus proches voisins de R a des temps d'exécution rédhibitoires sur ces données alors que l'implémentation de {\it Scikit-learn} est lente mais raisonnable. Comme cet algorithme ne s'adapte pas au le passage à l'échelle, il est abandonné les comparaisons pour se focaliser sur celui des \emph{forêts aléatoires} (\citet{brei-2001}). 

Les implémentations de R et {\it Scikit-learn} de ce dernier algorithme sont très voisines en terme de paramétrages et de valeurs par défaut.  Des forêts de 250 arbres sont comparées. En prendre plus n'améliore pas notablement la qualité de prévision, 200 suffisent en fait. Les autres valeurs par défaut des paramètres, notamment le nombre (racine carrée du nombre de pixels) de variables tirées aléatoirement : {\tt mtry} de {\it ranger}, {\tt max\_features} de {\it Scikit-learn},  {\tt featureSubsetStrategy} de {\it MLlib},  sont satisfaisantes. 

Pour satisfaire aux contraintes de \emph{MapReduce} l'implémentation de {\it MLlib} est basée sur un algorithme de construction d'arbres issu du projet Google PLANET (\citet{pand-2009}). Ce dernier dispose d'un nouveau paramètre : {\tt maxBins} ($=32$ par défaut) qui contrôle le nombre maximal de divisions prises en compte pour la recherche d'un n\oe ud optimal. Ce paramètre provoque le regroupement des modalités d'une variable qualitative qui en comporte trop ou la transformation d'une variable quantitative en une variable ordinale. Il permet de contrôler les temps d'exécution de la construction de chaque arbre et leur complexité. Les données MNIST étant finalement assez "binaires" (présence ou non d'un pixel), ce paramètre n'a quasiment aucun effet sur ces données à moins de lui donner une valeur très faible. 

Cette implémentation de \emph{random forest} dans {\it MLlib}, soulève des problèmes concrets plus délicats. Alors qu'en principe pour réduire le biais, les arbres sont estimés sans limitation ni élagage, des problèmes de gestion de mémoire liés à la configuration matérielle, imposent de contrôler la profondeur maximum des arbres ou ({\tt maxDepth}). En cas de manque de mémoire, Python (sous Windows, Mac OS ou Linux) gère ce problème par une extension ({\it swapping}) de ma mémoire, certes coûteuse en temps de calcul, sur le disque. En revanche ce même problème provoque une erreur et un arrêt intempestif de {\it Spark}. C'est donc par essais / erreurs successifs qu'il est possible de déterminer les valeurs des paramètres : nombre et profondeurs des arbres, acceptables sans erreur par chacun des n\oe uds du cluster. Des forêts de cent arbres de profondeur {\tt maxDetph = 15} sont estimées. 

\begin{figure}
\centerline{\includegraphics[width=12cm]{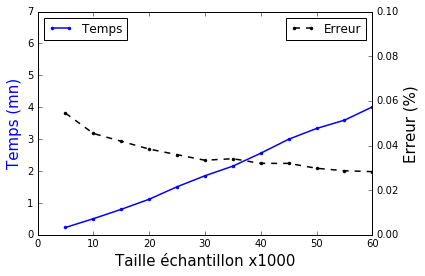}}
\caption{\it MNIST : forêts aléatoires avec R ({\it ranger}). \'Evolution du temps d'apprentissage et de l'erreur estimées sur l'échantillon test en fonction de la taille de l'échantillon d'apprentissage} \label{mnistResR}
\end{figure}

\begin{figure}
\centerline{\includegraphics[width=12cm]{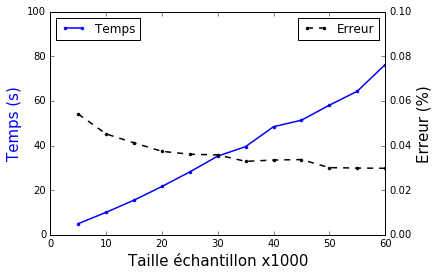}}
\caption{\it MNIST : forêts aléatoires avec Python ({\it Scikit-learn}). \'Evolution du temps d'apprentissage et de l'erreur estimées sur l'échantillon test en fonction de la taille de l'échantillon d'apprentissage.} \label{mnistResP}
\end{figure}

\begin{figure}
\centerline{\includegraphics[width=12cm]{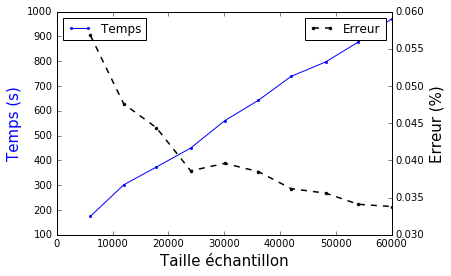}}
\caption{\it MNIST : forêts aléatoires avec Spark ({\it MLlib}). \'Evolution du temps d'apprentissage et de l'erreur estimées sur l'échantillon test en fonction de la taille de l'échantillon d'apprentissage.} \label{mnistResS}
\end{figure}

\begin{figure}
\centerline{\includegraphics[width=12cm]{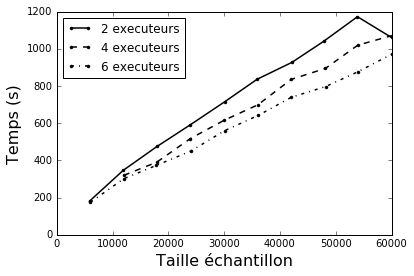}}
\caption{\it MNIST : Forêts aléatoires avec Spark ({\it MLlib}). \'Evolution du temps d'apprentissage pour plusieurs versions du cluster Spark avec 2, 4 ou 6 exécuteurs.} \label{mnistSpark}
\end{figure}

\subsection{Discussion}
La précision des prévisions entre R et Python est analogue (cf. figures~\ref{mnistResR}, \ref{mnistResP}), en revanche, les temps d'exécution, dépendent fortement du choix de la librairie R. Celle historique ({\it randomForest}), et intégrant le programme fortran de Breiman et Cutler interfacé avec R par \citet{liaw-2002}, s'avère très lente (facteur 20) en comparaison à d'autres implémentations écrites en C. La dernière en date ({\it ranger} de \citet{wrig-2016}) concurrence l'implémentation en Python et Cython de {\it Scikit-learn}. La différence entre les deux, au profit de Pyhton, vient de ce que cette dernière implémentation sait gérer la parallélisation (4 c\oe urs pour l'ordinateur utilisé) en fixant le paramètre {\tt n\_jobs=-1}, même sous Windows, contrairement à {\it ranger} qui ne peut le faire que sous un système issu d'Unix.

Les contraintes dues à la gestion de la mémoire de l'environnement {\it Spark} rendent difficiles l'obtention de résultats strictement similaires en terme de qualité de prévision (cf. figure~\ref{mnistResS}). La profondeur maximum des arbres influence fortement la qualité des estimations pour l'implémentation de {\it MLlib}. Le réglage de ce paramètre dépend directement de l'espace mémoire alloué à chaque exécuteur au moment de la déclaration du {\it cluster} par la commande {\tt spark\_context} ; une valeur de mémoire trop faible engendre une erreur qui arrête brutalement l'exécution. Seuls 100 arbres de profondeur maximum 15 sont estimés afin d'être sûr d'une exécution sans erreur. C'est une remise en cause de la tolérance aux pannes visées par {\it Spark} dans la version utilisée qui permet certes de gérer des données volumineuses mais pas, avec l'algorithme proposé, de mémoriser le modèle si celui-ci est trop complexe (nombre d'arbres, profondeur).  Il aurait fallut disposer d'un {\it cluster} plus conséquent (plus de mémoire par exécuteur) pour obtenir des résultats comparables en précision et en temps. D'autre part, le temps d'exécution décroît bien avec le nombre d'exécuteurs (cf. figure \ref{mnistSpark}) mais pas avec le facteur attendu pour atteindre la {\it scalabilité} de l'algorithme.

Enfin, remarquons que la croissance du temps d'apprentissage est linéaire en fonction de la taille de l'échantillon mais qu'au delà d'une taille d'échantillon de 50000, 55000 la précision des résultats semble arriver à un plateau (cf. figure~\ref{mnistResP}), surtout avec un nombre (250) relativement élevé d'arbres. Il aurait sans doute été utile de tester un sous-échantillonnage {\it sans remise} plutôt que le {\it bootstrap} classique des forêts aléatoires. Ceci doit permettre en principe de construire des modèles d'arbres encore moins corrélés entre eux et donc de réduire la variance des prévisions mais cette option d'échantillonnage n'est pas prévue dans les implémentations utilisées. Sans cette dernière possibilité, considérer des échantillons d'apprentissage encore plus volumineux n'est pas une priorité pour améliorer la précision de ce problème de reconnaissance de caractères, ce sont d'autres types de modélisation, possédant des propriétés d'invariance, qu'il faut déployer. 

En résumé, les expérimentations élémentaires réalisées sur ces données montrent qu'une architecture intégrée avec 8Go de mémoire et exécutant les implémentations {\it Python Scikit-learn} ou {\it R ranger} de {\it random forest} donne finalement de meilleurs résultats (temps, précision) qu'une architecture distribuée de 6 fois 7Go.

\section{Recommandation de films}
\subsection{Marketing et systèmes de recommandation}
\subsubsection*{Commerce en ligne}
La rapide expansion des sites de commerce en ligne a pour conséquence une explosion des besoins en marketing quantitatif et \emph{gestion de la relation client} (GRC) spécifiques à ce type de commerce. Ce domaine d'application est le principal moteur de développement des technologies liées au traitement des données massives et le premier fournisseur d'exemples. La GRC en marketing quantitatif traditionnel est principalement basée sur la construction de modèles de scores : d'appétence pour un produit, d'attrition (\emph{churn}) ou risque de rompre un contrat. Le commerce en ligne introduit de nouveaux enjeux avec le marché considérable de la publicité en ligne ou \emph{recommandation}.

\subsubsection*{Systèmes de recommandation}
La sélection ou recommandation automatique d'articles dans des approcches  appelées encore \emph{systèmes de recommandation} fait appel à des algorithmes dit de {\it filtrage}. Certains concernent des méthodes \emph{adaptatives} qui suivent la navigation de l'internaute, son flux de clics, jusqu'à l'achat ou non. Ces approches sont basées sur des algorithmes de bandits manchots et ne font pas l'objet de cet exemple. D'autres stratégies sont définies à partir d'un historique des comportements des clients, d'informations complémentaires sur leur profil, elles rejoignent les méthodes traditionnelles de marketing quantitatif. D'autres enfin sont basées sur la seule connaissance des interactions clients $\times$ produits à savoir la présence / absence d'achats ou un ensemble d'avis recueillis sous la forme de notes d'appréciation de chaque produit consommé. On parle alors de filtrage collaboratif (\emph{collaborative filtering}).

\subsubsection*{Filtrage collaboratif}
Ce dernier cas a largement été popularisé par le concours \emph{Netflix} o\`u il s'agit de proposer un film à un client en considérant seulement la matrice très creuse $\bm{X}$ : clients $\times$ films, des notes sur une échelle de 1 à 5.  L'objectif est donc de prévoir le goût, la note, ou l'appétence d'un client pour un produit (livre, film...), qu'il n'a pas acheté, afin de lui proposer celui le plus susceptible de répondre à ses attentes. 

Le filtrage collaboratif basé sur les seules interactions client $\times$ produits : présence / absence d'achat ou note d'appréciation  fait généralement appel à deux grandes familles de méthodes :
\begin{description}
\item[\it Méthodes de voisinage] fondées sur des indices de similarité (corrélation linéaire ou des rangs de Spearman...) entre clients ou (exclusif) entre produits :
\begin{itemize}
\item Basée sur le client avec l'hypothèse que des clients qui ont des préférences similaires vont apprécier des produits de façon similaire. Trouver un sous-ensemble $S_i$ de clients qui notent de manière similaire au client $i$ ; prévoir la note manquante : produit $j$ par client $i$,  par combinaison linéaire des notes de ce sous-ensemble $S_i$ sur le produit $j$. 
\item Basée sur le produit avec l'hypothèse que les clients préfèreront des produits similaires à ceux qu'ils ont déjà bien notés. Trouver un sous-ensemble $S_j$ de produits notés de façon similaire par le client $i$ ; prévoir la note manquante : produit $j$ par client $i$, par combinaison linéaire des notes de ce sous-ensemble $S_j$ du client $i$.  
\end{itemize}
\item[\it Modèle à facteurs latents] basé sur une décomposition de faible rang avec une éventuelle contrainte de régularisation, de la matrice très creuse $\bm{X}$ des notes ou avis clients $\times$ produits. La note du client $i$ sur le produit $j$ est approchée par le produit scalaire de deux facteurs $\bm{W}_i$ et $\bm{H}_j$ issus de la décomposition. 
\item[\it Complétion de matrice]. \`A la suite du succès du concours {\it Netflix}, \citet{cand-2010} ont formalisé de façon plus générale le problème de la recherche de facteurs latents comme celui de la complétion d'une matrice. Notons $P_\Omega(\bm{X})$ la "projection" de la matrice $\bm{X}$ qui consiste à remplacer toutes les valeurs inconnues (absence de note) par des valeurs nulles. 
\begin{equation}\label{compMat}
\min_{\bm{M}}{\left( ||P_\Omega(\bm{X}-\bm{M})||^2_2 + \lambda||\bm{M}||_* \right)}
\end{equation}
o\`u $||.||_2$ désigne la norme de Froebenius et $||.||_*$ celle nucléaire (somme des valeurs singulières) qui introduit une pénalisation afin de réduire le rang mais avec des propriétés de convexité. Le rôle de $P_\Omega(\bm{X})$ est important, l'ajustement n'est calculé que sur les seules valeurs connues de $\bm{X}$.
\end{description}
La littérature est très abondante sur le sujet qui soulève plusieurs questions dont :
\begin{itemize}
\item l'évaluation d'un système de recommandation car il n'est pas possible de savoir si le client a acheté sous l'effet de la proposition commerciale. Seul le montage complexe d'expérimentations (dites {\it AB testing}) permet de comparer deux stratégies sous forme de deux sites distincts auxquels sont aléatoirement adressés les internautes. 
\item Initialisation  (\emph{cold start problem}) des termes de la matrice avec l'arrivée de nouveaux clients ou la proposition de nouveaux produits.
\end{itemize}
Par ailleurs, des systèmes hybrides intègrent ces données d'interaction avec d'autres informations sur le profil des clients (âge, sexe, prénom...) ou encore sur la typologie des produits (genre, année...).

\subsection{Complétion de matrices}
La littérature sur ce sujet a connu un développement trop important pour espérer en faire une revue synthétique en quelques lignes. Nous nous limitons volontairement aux seules méthodes implémentées dans des librairies facilement accessibles des environnements R, Python et {\it Spark} considérés.
\subsubsection*{\it sofImpute}
\citet{mazu-2010} puis \citet{hast-2015} proposent deux algorithmes de complétions implémentés dans la librairie {\it softImpute} de R pour résoudre le problème d'optimisation (\ref{compMat}). 

Le premier (\citet{mazu-2010}) itère des décompositions en valeurs singulières (SVD) seuillées pour reconstruire les valeurs manquantes. Le seuillage de la SVD consiste à annuler les valeurs singulières de la décomposition inférieure à une valeur $\lambda$ forçant ainsi des solutions de plus faible rang. Des astuces permettent de réduire les temps de calcul. Il n'est pas nécessaire  de calculer tous les termes de la SVD et la faible densité des matrices permet des économies substantielles de mémoire. D'autre part les SVD successives peuvent bénéficier de l'itération précédente comme initialisation. 

Le deuxième algorithme  est basé sur une MMMF (maximum margin matrix factorization) de \citet{sreb-2005}. La matrice $\bm{X}$ à compléter est de dimension $(n\times p)$. Le critère à optimiser est donc 
$$
\min_{\bm{A_{n\times r},B_{p\times r}}}{||P_\Omega(\bm{X}-\bm{AB}^{T})||^2_2 + \lambda\left(||\bm{A}||^2_2+||\bm{B}||^2_2\right)},
$$
o\`u $r < \min(n,p)$ désigne le rang des matrices $\bm{A}$ et $\bm{B}$. Le critère n'est pas convexe mais dont la minimisation est résolue au moyen d'un algorithme de moindres carrés alternés (ALS). La minimisation est recherchée itérativement et alternativement en $\bm{A}$ puis $\bm{B}$. Si $\bm{A}$ est fixée, chercher la meilleure matrice $\bm{B}$ revient à résoudre $p$ régressions {\it ridge} : chaque colonne $\bm{X}_j$ est expliquée par les $r$ colonnes de $\bm{A}$ et pour les valeurs inconnues de $\bm{X}_j$, les lignes correspondantes de la matrice $\bm{A}$ sont ignorées de la $j$ème régression. La procédure équivalente est mise en place en fixant $\bm{B}$ et calculant $n$ régressions {\it ridge}. La d\'ecomposition obtenue sous la forme $\bm{X}=\bm{AB}^{T}$ est donc une approximation \`a partir des seules valeurs observ\'ees comportant des valeurs manquantes. Ces valeurs \`a compl\'eter sont obtenues alors par le r\'esultat du produit $\bm{AB}^{T}$. Bien entendu cela suppose que les valeurs manquantes ne sont pas des colonnes ou des lignes enti\`eres.

L'algorithme proposé par \citet{hast-2015} hybride les deux précédents : moindres carrés alternés et décomposition en valeurs singulières. Les expérimentations sur des données simulées et celles de {\it MovieLens} montrent une convergence plus rapide de la fonction objectif. 

Malheureusement, les auteurs ne donnent pas de résultats (RMSE) sur la précisions de la recommandation pour les données {\it MovieLens} et ceux fournis pour les données {\it Netflix} ne sont pas très convaincants. Une amélioration du score de base ({\it baseline}) de 0,1\%, à comparer avec les 10\% obtenus par la solution gagnante du concours. 

\subsubsection*{Complétion par NMF}
L'option {\tt ALS} de {\it softImpute} approche finalement le problème de complétion par une factorisation. Malheureusement,  \citet{hast-2015} n'introduisent pas dans leur comparaison, ni dans leurs références, d'autres approches de factorisation bien que la NMF (non negative matrix factorisation) soit largement citée et utilisée en filtrage collaboratif pour la recherche de facteurs latents sous la contrainte que toutes les valeurs soient positives ou nulles.
\paragraph{Problème d'optimisation.}
Comme la décomposition en valeurs singulières (SVD), la factorisation non négative de matrice (NMF) décompose une matrice rectangulaire $(n\times p)$ en le produit de deux matrices de faible rang. La solution de la SVD est unique, obtenue par des algorithmes efficaces et génère des facteurs orthogonaux permettant des représentations graphiques. En revanche la NMF fournit des résultats mieux adaptés à des notes, des effectifs, mais plusieurs algorithmes sont en concurrence qui convergent vers des optimums locaux voire même peuvent se bloquer sur une solution sous-optimale sur le bord du cône.
 
Soit $\bm{X}$ une matrice $(n\times p)$ ne contenant que des valeurs non négatives et sans ligne ou colonne ne comportant que des $0$ ;  $r$ un entier choisi relativement petit devant $n$ et $p$.  La factorisation non-négative de la matrice $\bm{X}$ est la recherche de deux matrices $\bm{W}_{n\times r}$ et $\bm{H}_{r\times p}$ ne contenant que des valeurs positives ou nulles et dont le produit approche $\bm{X}$.  
$$
\bm{X}\approx\bm{WH}.
$$
Le choix du \emph{rang} de factorisation $r<<min(n,p)$ assure une réduction drastique de dimension et donc des représentations parcimonieuses. \'Evidemment, la qualité d'approximation dépend de la parcimonie de la matrice initiale. 

La factorisation est résolue par la recherche d'un optimum local au problème d'optimisation :
$$
\min_{\bm{W, H}\geq0}\left[L(\bm{X},\bm{WH})+P(\bm{W, H})\right].
$$
$L$ est une fonction perte mesurant la qualité d'approximation et $P$ une fonction de pénalisation optionnelle ; $L$ est généralement soit un critère de moindres carrés (LS ou norme de Frobenius des matrices ou norme trace), soit la divergence de Kullback-Leibler (KL) ; $P$ est une pénalisation optionnelle de régularisation utilisée pour forcer les propriétés recherchées des matrices $\bm{W}$ et $\bm{H}$ ; par exemple, la parcimonie des matrices ou la régularité des solutions dans le cas de données spectrales.

Non seulement la solution est locale car la fonction objectif n'est pas convexe en $\bm{W}$ et $\bm{H}$ mais en plus la solution n'est pas unique. Toute matrice $\bm{D}_{r\times r}$ non négative et inversible fournit des solutions équivalentes en terme d'ajustement : 
$$
\bm{X}\approx\bm{WDD^{-1}H}.
$$

\paragraph{Algorithmes.}
De nombreuses variantes algorithmiques et sur la forme des pénalisations ont été publiées et implémentées généralement en Matlab, parfois en C, quelques unes spécifiques en R ; \citet{berr-2007} proposent un tour d'horizon de certaines tandis que \citet{gauj-2010} en ont implémentées dans R pour rendre facilement possible la comparaison des résultats. Trois familles d'algorithmes sont généralement citées :
\begin{itemize}
	\item {\it Standard NMF algorithm with multiplicative update},
	\item {\it Alternate Least Square (ALS) algorithm}, 
	\item Descente du gradient.
\end{itemize}
Chacun de ces algorithmes peut par ailleurs être initialisé de différentes façons :
\begin{itemize}
	\item plusieurs initialisations aléatoires de $\bm{W}$ et $\bm{H}$, le meilleur ajustement est conservé, 
	\item {\it non-negative double singular value decomposition (NNSVD)}, 
	\item une classification ($k$-means) des lignes ou des colonnes, 
	\item parts positives de matrices issues d'une analyse en composantes indépendantes (ACI),
	\item ...
\end{itemize}
Entre le choix de la fonction objectif : fonction perte (LS ou KL) et l'éventuelle pénalisation ($L^1$, $L^2$, régularité), le choix de l'algorithme ou d'une de ses variantes, le choix de l'initialisation... cela fait beaucoup d'options à comparer, tester. Comme toujours avec une nouvelle méthode et la pression de publication, de très nombreuses variantes apparaissent avant qu'une sélection {\it naturelle} n'opère pour aboutir à des choix plus efficaces et consensuels d'options en fonction du type de données traitées. 

\citet{berr-2007} décrivent très brièvement les principes de ces différents algorithmes et commentent leurs propriétés : convergence, complexité.

L'algorithme initial de \citet{lee-1999} (\emph{Multiplicative update algorithms}) peut converger vers un point stationnaire pas nécessairement minimum local, voire un point de la frontière même pas point stationnaire. Ces cas sont heureusement rares en pratique mais la convergence est considérée comme lente, demandant plus d'itérations que ses concurrents alors que chaque itération nécessite de nombreux calculs ($O(n^3)$). Les algorithmes de descente du gradient posent des questions délicates concernant le choix des deux pas de descente. La dernière famille d'algorithmes : moindres carrés alternés (ALS), exploite le fait que si le problème n'est pas à la fois convexe en $\bm{W}$ et $\bm{H}$, il l'est soit en $\bm{W}$ soit en $\bm{H}$. Il suit le principe ci-dessous et possède de bonnes propriétés (convergence, complexité). 

\begin{quote}
\noindent {\bf NMF par ALS}
 
\begin{itemize} 
\item $\bm{W}=$random$(n,r)$
\item {\bf FOR} $i=1$ à {\tt Maxiter}
\begin{itemize} 
\item Résoudre en $\bm{H}$ : $\bm{W'WH=W'X}$
\item Mettre à $0$ les termes négatifs de $\bm{H}$
\item Résoudre en $\bm{W}$ : $\bm{HH'W'=HX'}$
\item  Mettre à $0$ les termes négatifs de $\bm{W}$
\end{itemize}
\end{itemize}
\end{quote}

L'un des inconvénients du (\emph{Multiplicative update algorithms}) originel est que si un élément des matrices $\bm{W}$ ou $\bm{H}$ prend la valeur $0$, il reste à cette valeur, n'explorant ainsi pas de solutions alternatives. L'ALS est lui plus souple en permettant d'échapper à de mauvaises solutions locales.

\paragraph{Critères de choix.}
Les auteurs proposent différents critères pour aider aux choix des méthodes, algorithmes et paramètres, notamment celui du rang $r$ de factorisation, pouvant intervenir au cours d'une étude. L'évaluation de la ``stabilité'' de plusieurs exécutions de NMF repose sur des critères (silhouette, consensus, corrélation cophénétique) issues des méthodes de classification non supervisée. Pour adapter ces critères à la NMF, la notion de classe d'une observation (resp. d'une variable) est remplacée par la recherche du facteur, ou élément de la base (colonne de $\bm{W}$ resp. de $\bm{H}$), pour laquelle l'observation (resp. la variable) a obtenu la plus forte contribution. Comme pour le choix d'une dimension, d'un nombre de classes, seules des heuristiques sont proposées dans la littérature pour le difficile choix de $r$ pour lequel il n'y a pas de critère nettement tranché. 

En revanche, dans l'exemple traité de recherche d'une recommandation par facteurs latents, une approche de type supervisée est possible. Il suffit d'extraire de la matrice initiale un sous-ensemble de notes, nombre d'achats, qui constituera un sous-échantillon de validation pour optimiser le rang des matrices et l'éventuel paramètre de pénalisation. Plusieurs méthodes, plusieurs approches peuvent également être comparées sur un échantillon test construit de la même façon. Le concours \emph{Netflix} était construit sur ce principe avec une fonction perte quadratique (RMSE).

\paragraph{Implémentations.}
La librairie {\it NMF} de R implémente 11 méthodes ; 9 sont basées sur l'algorithme initial de \cite{lee-1999} (\emph{Multiplicative update algorithms}) avec différentes options de perte (LS, KL) et  de pénalisation ou d'arrêt, deux sont basées sur les moindres carrés alternés (ALS) avec contrainte de parcimonie sur les lignes ou les colonnes. Systématiquement, l'option est offerte, et encouragée, de lancer plusieurs exécutions à partir de plusieurs initialisations aléatoires pour sélectionner les options ``optimales'' puis, une fois les choix opérés,  pour retenir la meilleure parmi un ensemble d'exécutions.

Un algorithme spécifique (projection du gradient) et principalement utilisé pour de la décomposition d'images par NMF est accessible dans la librairie {\it Scikit-learn} pour matrices denses ou creuses. 

{\it Attention}, ces algorithmes  ne sont pas adaptés à la complétion de matrice. Appliqués à une matrice creuse de notes, ils visent à reconstruire précisément les valeurs manquantes considérées comme nulles et n'atteignent pas l'objectif recherché de compléter les valeurs manquantes.

{\it MLlib} implémente l'algorithme NMF par ALS avec deux options.  Dans celle par défaut, la fonction objectif est une norme $L_2$ (Froebenius) entre la matrice et sa reconstruction mais cette norme, comme dans le problème (\ref{compMat}) de complétion, n'est calculée \emph{que} sur les valeurs (notes) observées. Des contraintes ($L_1$ et $L_2$) introduisent une régularisation pour renforcer la parcimonie de la factorisation qui opère donc une complétion. Les paramètres sont à optimiser par validation croisée. La documentation n'est pas explicite sur le déroulement précis de l'algorithme mais cite \citet{kore-2009} qui le sont un peu plus. La deuxième option concerne des effectifs de ventes, de chargements de films. Les valeurs nulles ne sont pas des données manquantes.

\subsection{MovieLens}
\subsubsection*{Les données}
Des données réalistes croisant plusieurs milliers de clients et films, sont accessibles en ligne. Il s'agit d'une extraction du site {\tt movielens.org} qui vous aide à choisir un film. Quatre tailles de matrices très creuses sont proposées contenant seulement les appréciations connues d'un client sur un film. 
\begin{description}
\item[\tt 100k] 100 000 évaluations de 1000 utilisateurs de 1700 films.
\item[\tt 1M] Un million d'évaluations par 6000 utilisateurs sur 4000 films.
\item[\tt 10M] Dix millions d'évaluations par 72 000 utilisateurs sur 10 000 films.
\item[\tt 20M] Vingt deux millions d'évaluations par 138 000 utilisateurs sur 27 000 films. 
\end{description}

\subsubsection*{Complétion}
Les programmes ont été testés, sur la matrice de 100 000 notes puis directement appliqués à celle de 22 millions de notes. Les notes sont stockées dans une matrice de format creux ou {\it sparse}. Il s'agit simplement de mémoriser la liste des triplets contenant le numéro de la ligne (usager), de la colonne (film) pour lesquels la note est connue. La plus grande matrice se réfère donc à des dimensions de 138k lignes pour 27k colonnes ($3,7 10^9$ éléments) mais comme moins de 1\% des notes sont connues, elle tient en mémoire avec format creux.

L'évaluation de la complétion de matrice se ramène à un problème, ou tout du moins à la même démarche d'évaluation, qu'un problème d'apprentissage. La fonction perte utilisée pour mesure les reconstructions des notes est la racine de l'écart quadratique (RMSE). Un sous-ensemble de notes est extrait (10 \% soit près de deux millions de notes) pour jouer le rôle d'échantillon test. L'optimisation du rang de la factorisation et celle de la pénalisation peut être calculée par validation croisée mais compte tenu des tailles des matrices, l'extraction préalable d'un échantillon de validation est préférable. Seules deux approches ont été testées car {\it Scikit-learn} ne propose pas d'algorithme de complétion. 

La première approche a consisté à utiliser la librairie {\it softImpute} (option {\tt ALS}) de R pour plusieurs valeurs des paramètres (rang et pénalisation) afin de compléter l'échantillon test puis calculer le RMSE. Les résultats (processeur 4 c\oe urs séquencé à 1,7GHz, 8Go de RAM sous Windows) sont résumés dans le tableau~\ref{resSoftImpute}.

La deuxième approche met en \oe uvre la NMF de la librairie {\it MLlib}. Une procédure sommaire de validation croisée sur les données 100k a montré qu'un faible rang, entre 4 et 8 était préférable alors que le paramètre de pénalisation ($L_1$) est simplement laissé à sa valeur par défaut (0.01).

\subsubsection*{Discussion}

\begin{table}
\begin{center}
\caption{\it {\it SoftImpute} (ALS) appliqué aux données MovieLens. Temps (minutes) de calcul et RMSE en fonction du rang maximum de la factorisation et du paramètre de régularisation.}\label{resSoftImpute}
\vspace*{3mm}
\begin{tabular}{|crrr|}\hline
Rang Max & $\lambda$ & Temps & RMSE \\ \hline
 4  &  1       &  5.6 &  1.07   \\
 10 &  10  &  12.6 &  1.020  \\
 10 &  20  &  12.2 &  1.033 \\
 15 &  10  &  19.4 &  1.016 \\
 20 &   1  &  26.9  & 1.020 \\
 20 &  10  &  26.1  & 1.016 \\
 20 &  15  &  24.4 &  1.018 \\
 20 &  20  &  27.0  & 1.016  \\
 30 &  20  &  40.1 &  1.020 \\ \hline
\end{tabular}
\end{center}
\end{table}

Même avec une démarche qui pourrait conduire à du sur-apprentissage, la librairie de R {\it softImpute} conduit à des résultats décevants sur les données {\it MovieLens} (cf. tableau~\ref{resSoftImpute}) alors que la fonction NMF de {\it MLlib} conduit, sans effort d'optimisation élaboré, à des résultats satisfaisants (RMSE de 0,82), parmi les meilleurs trouvés dans les blogs de discussion sur le sujet.

\begin{figure}
\centerline{\includegraphics[width=12cm]{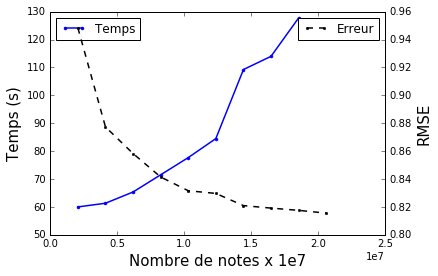}}
\caption{\it MovieLens : complétion par NMF ({\it MLlib}). \'Evolution du temps d'exécution de la factorisation et de l'erreur de complétion (RMSE) de l'échantillon test de notes en fonction du nombre de notes (taille de la matrice creuse) prises en compte dans la factorisation. }\label{mlensResS}
\end{figure}

\begin{figure}
\centerline{\includegraphics[width=12cm]{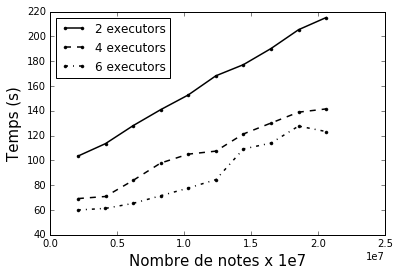}}
\caption{\it MovieLens : complétion par NMF ({\it MLlib}). \'Evolution du temps d'exécution de la factorisation en fonction du nombre de notes (taille de la matrice creuse) pour plusieurs versions du cluster avec 2, 4 ou 6 exécuteurs.} \label{mlensSpark}
\end{figure}

La figure~\ref{mlensResS} représente le RMSE et le temps de calcul pour la factorisation en fonction du nombre de notes prises en compte dans la matrice concernée. Il apparaît que le temps de calcul croît approximativement un peu plus vite que linéairement avec ce nombre de notes tandis que l'erreur continue de décroître mais de moins en moins rapidement. 

La figure~\ref{mlensSpark} montre par ailleurs que le temps de factorisation décroît avec le nombre d'exécuteurs du {\it cluster}. Ce ne sont pas les facteurs attendus pour une décroissance linéaire du temps en fonction de ce nombre d'exécuteurs mais le gain est significatif.

En résumé, la librairie {\it Scikit-learn} ne propose pas d'algorithme de complétion et est absente de la comparaison alors que {\it Spark MLlib} conduit aux meilleurs résultats, bien meilleurs en temps et précision que l'implémentation réalisée dans {\it softImpute} de R. Parmi une littérature déjà importante, ce n'est sans doute pas le meilleur algorithme qui est implémenté dans {\it MLlib} mais celui-ci (NMF par ALS) s'avère efficace et surtout bien adapté à l'architecture distribuée en association à {\it Spark} pour archiver, manipuler des matrices très creuses. 

\section{Catégorisation de produits}
\subsection{Objectifs}
\subsubsection*{Fouille de texte}
Il s'agit d'un  problème récurrent du commerce en ligne qui se présente sous la forme suivante. Un commerçant partenaire d'un site en ligne souhaite proposer l'ensemble d'un catalogue d'articles à la vente, chacun décrit par un court texte en langage naturel. Pour assurer le maximum de visibilité des produits, ce site doit assurer une catégorisation homogène des produits malgré leurs origines très variées : les commerçants partenaires. A partir de la liste des articles déjà présents sur le site (base d'apprentissage), il s'agit de déterminer la catégorie d'un nouvel article, c'est-à-dire permettre de l'introduire dans l'arborescence des catégories et sous-catégories du site ; c'est donc encore d'un problème de discrimination  ou classification supervisée mais appliquée à de la fouille de textes.

\subsubsection*{\'Etapes}
Le traitement se décompose en trois étapes bien distinctes. La première est un pré-traitement ou nettoyage des données. La deuxième est une  {\it vectorisation} ou quantification des textes, de façon à remplacer les mots par des nombres d'occurrences ou plutôt par les valeurs prises par une liste de variables ({\it features}) mesurant des fréquences relatives d'une liste déterminée de regroupements de mots. Enfin, une fois construite une matrice, généralement  creuse, différentes méthodes d'apprentissage sont testées dans la 3ème étape afin de prévoir, au mieux, la catégorie des articles d'un échantillon test.

La préparation des données ({\it data munging}) est très souvent l'étape la plus délicate et la plus {\it chronophage} de la Science des Données de la vraie vie, par opposition à des données rendues publiques et souvent exploitées pour illustrer ou comparer des méthodes. Ces dernières, comme celles de l'exemple de reconnaissance des caractères, sont très \og propres \fg{} : pas de données manquantes, ou trop atypiques, d'erreur de codage...  Souvent négligé des présentations pédagogiques, le pré-traitement est néanmoins primordial pour assurer la qualité et le pouvoir prédictif des nouvelles variables ainsi construites et qui influent directement sur la pertinence des modèles. D'autre part, ces phases d'extraction, nettoyage, recodage, transformation... des données, conduisent très souvent à une réduction drastique de leur volume. Des données initialement massives, il ressort une matrice souvent adaptée à la mémoire d'un plus ou moins gros ordinateur et ces pré-traitements peuvent rendre inutile une architecture distribuée pour la suite des analyses.

\subsection{Préparation des textes}
Voici la liste des traitements généralement opérés sur des données textuelles.
\begin{description}
\item[\it Nettoyage] Suppression des caractères mal codés et de ponctuation, transformation des majuscules en minuscules, en remarquant que ces transformations ne seraient pas pertinentes pour un objectif de détection de pourriels.
\item[\it Suppression] des mots vides ({\it stop words}) ou mots de liaison, articles qui n'ont {\it a priori} pas de pouvoir discriminant.
\item[\it Racinisation] ou {\it stemming}. Les mots sont réduits à leur seule racine afin de réduire la taille du dictionnaire. 
\item[\it Hashage] Une fonction de hashage est appliquée pour transformer chaque mot en un index unique en ajoutant une {\it astuce} ou {\it hashing trick}. Le nombre de valeurs possibles (modulo une division entière) prises par la fonction de hashage  est un paramètre {\tt n\_hash}. Ainsi, les mots sont automatiquement et arbitrairement regroupés pour aboutir à un nombre prédéterminé de codes possibles. Plus précisément, la fonction de hashage $h$ est définie sur l'espace des entiers naturels et à valeurs $i=h(j)$ dans un ensemble fini $(1,\ldots, \mbox{\tt n\_hash})$ des variables ou {\it features}. Ainsi le poids de l'indice $i$, du nouvel espace, est l'association de tous les poids d'indice $j$ tels que $i=h(j)$ de l'espace original. Ici, les poids sont associés d'après la méthode décrite par \citet{wein-2009}. La fonction $h$ n'est pas générée aléatoirement. Ainsi pour un même fichier d'apprentissage (ou de test) et pour un même entier {\tt n\_hash}, le résultat de la fonction de hashage est identique. 
\item[\it Xgram] La fonction de hashage est appliquée aux mots ({\tt unigram}) ou aux couples ({\tt bigram}) de deux mots consécutifs. Ce deuxième choix permet de lever beaucoup d'ambiguïté du langage mais risque de faire exploser le volume du dictionnaire. C'est encore un paramètre à optimiser.
\item[\it TF-IDF] Il  permet de faire ressortir l'importance relative de chaque mot $m$ (ou couples de mots consécutifs) dans un texte-produit ou un document $d$, par rapport à la liste entière des documents. La fonction $TF(m,d)$ compte le nombre d'occurrences du mot $m$ dans le document $d$. La fonction $IDF(m)$  mesure l'importance du terme dans l'ensemble des documents ou descriptifs en donnant plus de poids aux termes les moins fréquents car considérés comme les plus discriminants (motivation analogue à celle de la métrique du $\chi$2 en analyse des correspondances). $IDF(m)=\log\frac{D+1}{f(m)+1}$ (version {\it smooth} adoptée dans {\it Scikit-learn} et {\it MLlib}) o\`u $D$ est le nombre de documents, la taille de l'échantillon d'apprentissage, et $f(m)$ le nombre de documents ou descriptifs contenant le mot $m$. La nouvelle variable ou {\it features} est $V_m(d)=TF(m,d)\times IDF(m)$. 
\end{description}
Comme pour les transformations des variables quantitatives (centrage, réduction), les mêmes transformations, c'est-à-dire la même fonction de hashage et le même ensemble de pondérations (IDF), sont calculés, appliqués sur l'échantillon d'apprentissage puis appliqués sur celui de test. 

Il s'agit finalement d'évaluer distinctement ces trois étapes au regard des technologies disponibles. De façon schématique, la première (nettoyage) est très facilement parallélisable et peut se décomposer en séquences de traitements ou d'étapes fonctionnelles {\it Map} tout à fait adaptées à une architecture distribuée et donc des données très massives. En revanche, les étapes suivantes (vectorisation, modélisation) nécessitent des comparaisons plus fouillées.

\subsection{\it Cdiscount}
Il s'agit d'une version simplifiée du concours proposé par {\it Cdiscount} et paru sur le site {\tt datascience.net}. Les données d'apprentissage sont accessibles sur demande auprès de {\it Cdiscount} dans le forum de ce site et les solutions gagnantes ont été présentées aux journées de Statistique de Montpellier par \citet{gout-2016}. Comme les solutions de l'échantillon test du concours ne sont pas et ne seront pas rendues publiques, un échantillon test est donc extrait pour l'usage de cet exemple. L'objectif est de prévoir la catégorie d'un produit à partir de son descriptif. Seule la catégorie principale (1er niveau de 47 classes) est modélisée au lieu des trois niveaux demandés dans le concours (5789 classes). L'objectif n'est pas de faire mieux que les solutions gagnantes basées sur des {\it pyramides} complexes de régressions logistiques programmées en Python mais de comparer les performances des méthodes et technologies en fonction de la taille de la base d'apprentissage ainsi que d'illustrer, sur un exemple réel, le pré-traitement de données textuelles. La stratégie de sous ou sur-échantillonnage des catégories très déséquilibrées qui permet d'améliorer la prévision n'a pas été mise en \oe uvre.

Les données se présentent sous la forme d'un fichier texte de 3.5 Go. Il comporte quinze millions de lignes, un produit par ligne contenant sa catégorie et le descriptif. La nature brute de ces données, leur volume, permet de considérer toute la chaîne de traitement et pas seulement la partie apprentissage.

R, peu adapté à une fouille de textes d'un tel volume, n'a pas été testé, seules sont comparées deux séquences d'analyses identiques réalisées avec Python {\it Scikit-learn} et {\it Spark Mllib}.

\subsubsection*{Nettoyage}
La première étape opère une série de traitements élémentaires (suppression des caractères mal codés et de ponctuation, transformation des majuscules en minuscules) puis certains adaptés à l'objectif : suppression des mots vides ({\it stop words}) et {\it racinisation} ou {\it stemming} des mots.
\begin{figure}
\centerline{\includegraphics[width=12cm]{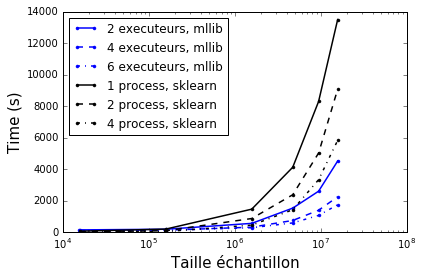}}
\caption{\it Cdiscount : nettoyage des données. \'Evolution du temps de nettoyage pour plusieurs versions du cluster Spark avec 2, 4 ou 6 exécuteurs et avec Scikit-learnet en Python (Scikit-learn) avec 1, 2, 4 processeurs.} \label{cdiscountClean}
\end{figure}

Ces traitements ont été exécutés avec différentes configurations du {\it cluster} (2, 4, 6 exécuteurs de 7Go) ainsi qu'en contrôlant le nombre de c\oe urs (1,2 ou 4) actifs d'un Macbook Pro (2,2 GHz) avec 16Go de RAM. L'échelle des abscisses (figure~\ref{cdiscountClean}) est logarithmique mais le temps d'exécution croît approximativement linéairement avec le volume des données à nettoyer. Les résultats de {\emph scalabilité} sont ceux attendus en fonction du nombre de c\oe urs ou du nombre d'exécuteurs et nettement en faveur de l'architecture distribuée. Intuitivement, une architecture physiquement plutôt que virtuellement distribuée devrait encore conduire à de meilleurs résultats sur de très gros volumes avec une parallélisation plus efficace des opérations de lecture. 

En résumé, une architecture distribuée avec {\it Spark} est plus efficace pour l'étape de nettoyage qu'une architecture intégrée.

\subsubsection*{\it Vectorisation}
\begin{figure}
\centerline{\includegraphics[width=12cm]{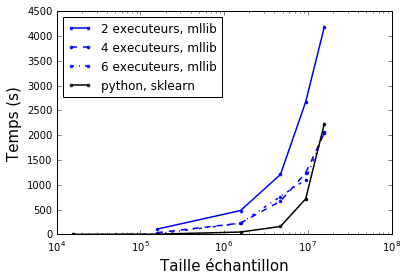}}
\caption{\it Cdiscount : vectorisation. \'Evolution du temps de la vectorisation (hashage et TF-IDF) pour plusieurs versions du cluster Spark avec 2, 4 ou 6 exécuteurs et en Python (Scikit-learn) avec {\tt n\_hash} = 60 000 variables.} \label{cdiscountHasidf}
\end{figure}

Les étapes de hashage puis de calcul des fréquences (TF) et des pondérations (IDF) sont nécessairement associées avec la gestion d'un dictionnaire commun dont le cardinal est la valeur du paramètre {\tt n\_hash}. Le résultat de cette étape est une matrice de fréquences relatives comportant autant de lignes que de textes et {\tt n\_hash} colonnes. Les documentations ne sont pas très explicites sur les implémentations de ces traitements mais, de toute façon, plusieurs étapes (au moins 3 ?) {\it Mapreduce} sont nécessaires avec {\it Spark MLlib}. Des valeurs plus grandes (100k, 500k, 1M) de {\tt n\_hash} ont été testées mais comme
\begin{itemize}
\item la précision (erreur de classement) obtenue avec {\it Scikit-learn} n'est pas meilleure au delà de {\tt n\_hash =60 000} variables,
\item {\it Spark Mllib} provoque une erreur mémoire à partir de {\tt n\_hash =100 000} variables dans l'étape suivante d'apprentissage,
\end{itemize}
seuls les résultats pour des valeurs plus faibles car utiles de {\tt n\_hash} (10k, 35k, 60k) sont considérés. D'autre part, les graphiques ne sont pas superposés par souci de lisibilité mais la valeur du paramètre {\tt n\_hash} n'influe pas très sensiblement sur le temps d'exécution.

La figure \ref{cdiscountHasidf} montre des temps d'exécution en faveur d'une architecture intégrée (Macbook pro,  16Go), à moins sans doute de traiter des données très volumineuses car dans ce cas, la gestion de la mémoire virtuelles({\it swapping}) pénalise l'exécution. Mais, il suffirait alors de renforcer la RAM au niveau, par exemple, de celle globale du {\it cluster} pour les temps concurrentiels.

En résumé, une architecture intégrée, en renforçant éventuellement la mémoire, est préférable à une architecture distribuée pour l'étape de {\it vectorisation} de documents.
\subsection*{Apprentissage}
Plusieurs méthodes d'apprentissage (arbres, {\it random forest}) ont été testées sur ces données pour finalement choisir une ou plutôt un ensemble de 47 régressions logistiques. En effet, par défaut, {\it Scikit-learn}, comme {\it MLlib}, estiment une version multimodale de la régression logistique en considérant des modèles prévoyant l'occurrence d'une classe contre celle des autres. Il est important de noter que, pour le problème initial, l'objectif est de prévoir le troisième niveau de catégorie qui comporte 5789 classes. Les solutions gagnantes, estimées avec des valeurs de {\tt n\_hash} = 500k, produisent donc un ensemble de modèles de régressions logistiques nécessitant 20Go de mémoire rien que pour l'archivage des paramètres. Bien sûr la pénalisation Lasso engendre des solutions creuses mais il n'est pas sûr que l'espace mémoire des paramètres mis à 0 soit bien économisé et il ne l'est de toute façon pas an début d'exécution.

Par ailleurs, les deux algorithmes d'optimisation ({\it LBGFS et Liblinear}) proposés pour estimer une régression logistique ont été comparés sans mettre en évidence de différence ; les résultats ne sont pas présentés.

\begin{figure}
\centerline{\includegraphics[width=12cm]{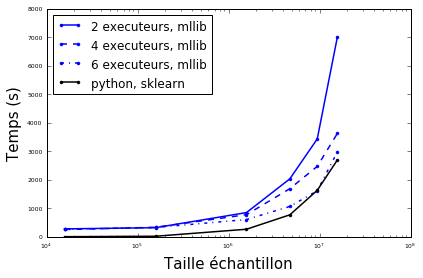}}
\caption{\it Cdiscount : apprentissage. \'Evolution du temps d'apprentissage pour plusieurs versions du cluster Spark avec 2, 4 ou 6 exécuteurs et en Python (Scikit-learn) avec {\tt n\_hash} = 60 000 variables.} \label{cdiscountApprent}
\end{figure}
la figure \ref{cdiscountApprent} compare les temps d'exécution pour l'estimation des modèles en fonction de la taille de l'échantillon. Le comportement est assez identique à celui de l'étape (vectorisation) précédente. L'architecture intégrée s'avère plus performante à condition de disposer de suffisamment de mémoire.

\begin{figure}
\centerline{\includegraphics[width=12cm]{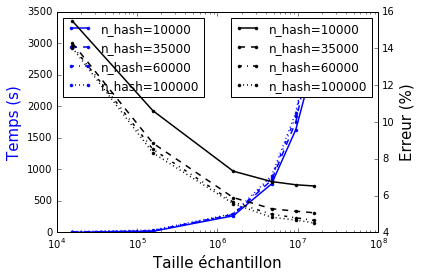}}
\caption{\it Cdiscount. \'Evolution du temps d'exécution de l'apprentissage avec Scikit-learn et du taux d'erreur sur l'échantillon test en fonction de la taille de l'échantillon  et du nombre {\tt n\_hash} de variables ou mots du dictionnaire. }\label{cdiscountResP}
\end{figure}
\begin{figure}
\centerline{\includegraphics[width=12cm]{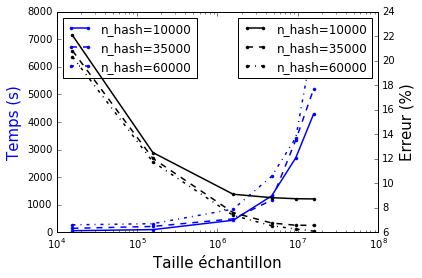}}
\caption{\it Cdiscount. \'Evolution du temps d'exécution de l'apprentissage avec Spark MLlib et du taux d'erreur sur l'échantillon test en fonction de la taille de l'échantillon et du nombre {\tt n\_hash} de variables ou mots du dictionnaire.}\label{cdiscountResS}
\end{figure}

Les figures \ref{cdiscountResP} et \ref{cdiscountResS} représentent les évolutions du taux d'erreur et du temps d'exécution de l'estimation des modèles en fonction de la taille de l'échantillon d'apprentissage et ce pour plusieurs valeurs du paramètre {\tt n\_hash}. La croissance du temps est plutôt linéaire avec la taille de l'échantillon tandis que ce temps est assez insensible au nombre de variables ({\tt n\_hash}) pour {\it Scikit-learn} contrairement à {\it MLlib}. 

Le taux d'erreur décroit avec la taille de l'échantillon d'apprentissage mais ne semble pas encore stabilisée avec 15M de produits. Il décroît également avec le nombre de variables mais se stabilise au delà de 60~000. Sans explication claire, les régressions logistiques estimées par {\it MLlib} stagnent à des taux d'erreur supérieurs à celles estimées par {\it Scikit-learn} même en limitant ces dernières au même nombre de variables. Enfin ajouter plus de variables (75k) pour {\it Spark MLlib} provoquent des erreurs d'exécution pour défaut de mémoire, contrairement à Python {\it Scikit-learn} qui accepte des valeurs de {\tt n\_hash} nettement plus grandes (1M) en précisant que cette fois ({\it swapping}) les temps d'exécution deviennent nettement plus importants.

En résumé, pour cette étape de modélisation par un ensemble de régressions logistiques, l'architecture intégrée se montre plus efficace en temps et en précision que l'architecture distribuée qui nécessite, globalement, des ressources mémoires importantes pour éviter des erreurs à l'exécution.

\section{Conclusion}

Finalement la comparaison s'établit principalement entre deux types d'architecture de parallélisation pour l'analyse de données massives ; celle \emph{intégrée} du calcul à haute performance avec un serveur opérant de nombreux processeurs et beaucoup de mémoire (RAM) ou celle des données et calculs {\it distribuées} sur des n\oe uds physiquement distincts. Pour ajouter à la confusion, les architectures "distribuées" testées le furent virtuellement avec la définition de {\it n\oe uds virtuels} au sein d'une architecture intégrée. C'est d'expérience très souvent le cas et ce qui est alors recherchée est plus la capacité de {\it Spark} à gérer des données hétérogènes, qu'une réelle parallélisation des entrées / sorties de données excessivement massives. 

La mise en \oe uvre de solutions {\it Hadoop Spark} ajoute une couche de complexité supplémentaire. Cet investissement n'est à tenter que mûrement réfléchi et justifié. Les comparaisons réalisées sur ces trois cas d'usage amènent quelques réflexions dont la validité reste limitée dans le temps avec le niveau de maturité de ces technologies  et même avec leur simple existence. Ainsi {\it Spark} dont une diffusion stable n'est disponible que depuis 2014 est en plein développement ; la version 2 est annoncée pour septembre 2016 de même que la version stable 0.18 de {\it Scikit-learn}. 

La pression académique de publication conduit à la production de méthodes ou variantes de méthodes dont les performances ne sont montrées que sur des exemples sélectionnés. Parallèlement, la pression commerciale engendre des battages médiatiques mettant exagérément en valeur les avantages de nouvelles technologies. Même limitées et sans doute assez naïves, les quelques expérimentations conduites ici permettent de relativiser certains jugements. 

\begin{itemize}
\item Pour la gestion ({\it munging}) de volumes et/ou de flux de données importants, complexes, hétérogènes, la parallélisation efficace de traitements élémentaires, les fonctionnalités déjà opérationnelles de {\it SparkSQL}, {\it streaming, pipeline} sont des atouts à prendre en compte.
\item En revanche, le recours à {\it MLlib} ou {\it SparkML} pour la phase d'apprentissage ou de modélisation (TF-IDF incluse) n'est pas, en l'état du développement de ces librairies, une priorité. Les librairies de Python ({\it Scikit-learn}), R, éventuellement celles en développement de Julia, répondent aux besoins en utilisant une machine avec suffisamment de mémoire et de processeurs. La distribution des algorithmes ({\it MapReduce}) sur plusieurs machines soulève plus de problèmes, notamment de gestion de la mémoire, qu'elle n'en résout pour des algorithmes complexes.
\item Il ne faut pas perdre de vue que des {\it big data} peuvent engendrer des {\it big} modèles de régression, de forêts aléatoires... sans même aller jusqu'à la complexité des solutions gagnantes de concours {\it Kaggle}. Il ne suffit plus de gérer les données, il faut aussi mémoriser ces modèles lors de leur estimation puis en exploitation. Une architecture intégrée de calcul haute performance (HPC) semble plus adaptée à l'utilisation de modèles complexes qu'une architecture distribuée.
\item Néanmoins, utiliser {\it SparkML, MLlib} en {\it modélisation} si la mise en \oe uvre d'un algorithme distribué reste simple, sans risque de dépassement mémoire ; c'est le cas des problèmes de recommandation par complétion de matrice.  L'algorithme de factorisation non négative (NMF) par moindre carrés alternés (ALS) {\it MLlib} se montre simple et efficace pour gérer des grandes matrices creuses et produire des matrices de facteurs latents de faible rang donc de volume accessible.
\item Ne pas perdre de vue que les comparaisons de cet article se limitent aux méthodes implémentées dans les libraires librement accessibles les plus en vogue. Il y a un fossé avec des développements académiques récents (\citet{genu-2015}) pas ou pas encore implémentés dans ces librairies. Le temps et une forme de sélection naturelle fera le tri des approches les plus efficaces afin de combler ce fossé.
\item Entre R et Python, le choix est assez facile et dépend de l'objectif, de la complexité des flots de traitement. Schématiquement, R propose beaucoup plus d'outils et de méthodes permettant de comprendre, interpréter, donner du sens aux modèles. En revanche Python se montre plus efficace dans les phases de traitements et exploitations, pour la prévision brute. Les deux environnements sont, en l'état actuel, plutôt complémentaires.
\end{itemize}

En définitive, les conclusions de ces expérimentations ne présentent rien d'extraordinaire ; elles sont fidèles à une forme de bon sens et correspondent aux pratiques industrielles présentées dans différents forums et conférences (Cdiscount, Critéo, Deepky, Tinyclues...) ; il est néanmoins utile de le vérifier concrètement, indépendamment de la pression commerciale des éditeurs de logiciels ou de celle, de publication, des chercheurs académiques. Une architecture distribuée ({\it Hadoop Spark}) est adaptée à la gestion de gros volumes et flux de données en ligne, alors que la modélisation ou l'apprentissage peut se faire hors ligne sur des sous-ensembles des données avec une architecture parallèle et une mémoire importante intégrée. Les besoins en calcul priment alors sur les entrées / sorties.  D'autres approches d'apprentissage en ligne, notamment par des algorithmes de gradient stochastique, soulèvent d'autres questions d'indépendance du flux de données.

Il est nécessaire de rappeler que la qualité de la {\it Science des Données} dépend directement de leur fiabilité ({\it garbage in garbage out}), de leur représentativité. L'exemple des données MNIST montre bien qu'au delà d'un seuil, la taille des données n'améliore pas la précision. En revanche, la {\it construction d'une base d'apprentissage} en insistant éventuellement sur les zones de l'espace plus difficiles à prédire, tout en allégeant celles plus faciles, reste un défi majeur à relever dans chaque cas étudié. Enfin ce travail n'aborde pas les algorithmes d'apprentissage profond nécessitant des moyens de calcul et bases d'apprentissage d'une autre nature pour faire le choix, par exemple en analyse d'images, entre construction de variables caractéristiques (\it features}) discriminantes puis apprentissage classique et analyse brutes des images ou signaux par {\it deep learning}.

\bibliographystyle{jesnat-fr-utf8}

\bibliography{biblio-utf8}

\end{document}